\documentclass[a4paper,11pt]{article}
\pdfoutput=1 

\usepackage{jcappub} 

\usepackage[T1]{fontenc} 

\title{\boldmath Exact Radiation Model For Perfect Fluid Under Maximum Entropy Principle}

\author[a]{Abdul Aziz,}
\author[a]{Sourav Roy Chowdhury,}
\author[a]{Debabrata Deb,}
\author[b]{Farook Rahaman,}
\author[c]{Saibal Ray,}
\author[a]{B.K. Guha.}

\affiliation[a]{Department of Physics, Indian Institute of Engineering Science and Technology, Shibpur, Howrah 711103, West Bengal, India}
\affiliation[b]{Department of Mathematics, Jadavpur University, Kolkata
700032, West Bengal, India}
\affiliation[c]{Department of Physics, Government College of
	Engineering and Ceramic Technology, Kolkata 700010, West Bengal,
	India}

\emailAdd{azizmail2012@gmail.com}
\emailAdd{sourav.rs2016@physics.iiests.ac.in}
\emailAdd{ddeb.rs2016@physics.iiests.ac.in}
\emailAdd{rahaman@associates.iucaa.in}
\emailAdd{saibal@associates.iucaa.in}
\emailAdd{bkguhaphys@gmail.com}

\abstract{We find the Euler-Lagrangian equation by maximising the total entropy. Hence we obtain an expression for mass of the spherically symmetric system by solving the Euler-Lagrangian equation where the Homotopy Perturbation Method has been employed. With the help of this expression and the Einstein field equations we obtain an interior solution set. Thereafter, we explain different aspects of the solution describing the system in connection to the mass, density, pressures, energy, stability, mass-radius ratio, compactness factor and surface redshift. This analysis shows that all the physical properties, in connection to brown dwarf stars, are valid with the observed features.}


\begin{document}
\maketitle
\flushbottom

\section{Introduction}
There has always been search for the interior solution of the spherically symmetric system. Scientists have tried to find out more than several hundreds different type of interior solutions. But very few solutions made its physical acceptance in all aspects describing the system so far.

However, in the present paper we have studied a spherically symmetric system of radiating star under the homotopy perturbation
method (HPM) which was introduced and developed by~\citep{He1997,He1999,He2000,He2004,He2005,He2006,He2010} and others~\citep{Liao1992,Liu1997,Liao2003,Liao2004,Liao2012}. This is a series expansion method used in the solution of non-linear  partial differential equations, in the present case the Einstein field equations of general relativity. The method in principle employs a homotopy transform to generate a convergent series solution of differential equations.	~\citet{He2000} advocated in favour of homotopy as well as perturbation technique to
solve non-linear problems. Subsequently other researchers also applied the HPM in various field of pure and applied mathematics (as an exhaustive appliance), and in physics and astrophysics (as a new field of application) to solve related non-linear differential equations in an extraordinary simplified way~\citep{Finkelstein2006,Demir2013,Siddiqi2013,Nourazar2015}.

In connection to self-gravitating radiation model~\citet{Sorkin1981} have examined the entropy of self-gravitating
radiation confined to a spherical box with finite radius in the context of general relativity. Their results are expected to
supports the validity of self-gravitating systems of the Bekenstein upper limit on the entropy to energy ratio of material bodies.

~\citet{Rahaman2014} proposed and analysed a model for the existence of strange stars. They predicted a mass function for the ultra dense strange stars. The interpolation technique has been used to estimate the cubic polynomial that yield the following expression for the mass as a function of the radial coordinate $m(r) = ar^3 - br^2 + cr - d$ with $a$, $b$, $c$ and $d$ all being numerical constants. Their analysis is based on the MIT bag model and yields physically valid energy density, radial and transverse pressures.

However, in the above mentioned work of~\citet{Rahaman2014} the target object was a strange star. In our
present investigation, we start with the intention to develop a basic interior solution of the Einstein equations valid for any radiating model under a similar expression for the mass as a function of the radial coordinate, i.e., $m(r)$. Then we match our theoretically obtained solution set with the observational results for practical validity of the model and find that our model is the best fit for the brown dwarf star of $E0$ type.

As a consequence, in the present study our sole motivation is to find the interior solution of the spherically symmetric system considering radiation effects in a more general way. For this we consider a metric with unknown time-time component. By using the maximum
entropy principle (MEP) we obtain an Euler-Lagrangian equation which gives second-order differential equation for mass of the
spherical distribution. Then we find out solution for mass by the homotopy perturbation method (HPM). The Einstein field equations
are solved for density, time-time component of metric and pressures of the system. Thus, the main purpose of the present investigation is to take data set
arbitrarily and see the physical nature and from there to predict whether this type of radiation model can effectively be used.

The work plan of the study is as follows: in section 2 we develop the methodology for the mass of the spherical symmetric system of
radiating star under (i) the maximum entropy principle, and (ii) the homotopy perturbation method. Next in the section 3 we provide
the Einstein field equations for self-gravitating radiation system and discuss several physical aspects of the solution describing
the system in connection to mass, density, pressures, energy, stability, mass-radius ratio, compactness factor and surface
redshift. In section 4 it has been shown in details that all the physical features in connection to brown dwarf stars of type $E0$
are valid with the observed features. We have made some specific remarks on the radiation model in the section 5.

\section{The methodology for mass of the spherical symmetric system of radiating star}

\subsection{The Maximum Entropy Principle (MEP)}
We consider the metric of the spherical system
\begin{equation}
ds^{2}= -g_{tt}(r)dt^{2}+\left(1-\frac{2m(r)}{r}\right)^{-1}dr^{2}+r^{2}(d\theta^{2}+sin^{2}\theta d\phi^{2}), \label{eq1}
\end{equation}
where $m(r)$ is the mass distribution of the spherically symmetric body and $g_{tt}(r)$ is the time-time component of
metric. These are functions of the radial coordinate $r$ only.

The equation of state of the system with thermal radiation is given by
\begin{equation}
p = \frac{1}{3}\rho, \label{eq2}
\end{equation}
where $p$ and $\rho$ are the radial pressure and density of the matter distribution. Compatible with spherically symmetry, we assume the general energy-momentum tensor in the interior of the isotropic star as
\begin{equation}
T_\nu^\mu=diag(\rho,-p,-p,-p). \label{eq:emten}
\end{equation}

For the locally measured temperature $T$ of the black body radiation, the rest frame energy density and entropy density, respectively, assume
the following forms~\citep{Sorkin1981}
\begin{equation}
\rho=bT^{4},\label{eq3a}
\end{equation}

\begin{equation}
s = \frac{4}{3}bT^{3}, \label{eq3b}
\end{equation}
where $b$ is a constant. Here we have assumed that in Planck units $G = c = h = k = 1$~\citep{Gibbons1978}.

By substituting eq.  (\ref{eq3a})  into eq.  (\ref{eq3b}), one can easily obtain the following relation
\begin{equation}
s=\alpha(\rho)^{3/4}, \label{eq4a}
\end{equation}
where $\alpha=\frac{4}{3}b^{1/4}$.

For the matter distribution up to radius $r \leq R$, the total entropy is given by
\begin{equation}
S = 4\pi\int_{0}^{R}s(r)\left[1-\frac{2m(r)}{r}\right]^{-1/2}r^{2}dr,
\label{eq5}
\end{equation}
which can be simplified as
\begin{equation}
S = (4\pi)^{\frac{1}{4}}\alpha\int_{0}^{R}Ldr, \label{eq6}
\end{equation}
where the Lagrangian $L$ is given by
\begin{equation}
L=(m^{\prime})^{\frac{3}{4}}\left[1-\frac{2m(r)}{r}\right]^{-\frac{1}{2}}r^{\frac{1}{2}}, \label{eq7}
\end{equation}
with constraint equation which we get from Einstein's field equation
\begin{equation}
m^\prime(r) = 4 \pi r^2 \rho. \label{eq7a}
\end{equation}

Also, we have the Euler-Lagrangian equation in the form
\begin{equation}
\frac{d}{dr}\left(\frac{\partial{L}}{\partial{m^{\prime}}}\right)
- \frac{\partial{L}}{\partial{m}} = 0, \label{eq8}
\end{equation}
which reduces to
\begin{equation}
m^{\prime} + m^{\prime\prime}m - \frac{1}{2}m^{\prime\prime}r -
\frac{2}{3}(m^{\prime})^{2} - \frac{4mm^{\prime}}{r} = 0. \label{eq9}
\end{equation}

Here the above eq. (\ref{eq9}) is the master equation for a self-gravitating radiation  system. If it is possible to solve this highly non-linear
differential equation, then one would get the complete structure of the self-gravitating radiation system and therefore could be used
to describe a model of a fluid sphere or stellar configuration. In the following section, we shall solve this equation by using a
newly developed technique known as {\it homotopy perturbation method} (HPM)~\citep{He2000}.

Substituting eq. (\ref{eq7a}) in eq. (\ref{eq9}), we obtain relativistic Tolman-Oppenheimer-Volkoff (TOV) equation 	
as follows:
 \begin{equation}
\frac{d}{dr} \left(\frac{\rho}{3}\right)=-\frac{4\rho}{9r}\left(\frac{3m+rm^{\prime}}{r-2m}\right).\label{TOV}
 \end{equation}
 
 The  relativistic Tolman-Oppenheimer-Volkoff (TOV) equation  for the isotropic stable system given in eq. (\ref{TOV}) can be derived from the Tolman-Whittaker formula and the Einstein field equations in the following explicit structure
\begin{equation}
-\frac{M_G\left(\rho+p\right)}{r^2}e^{\frac{\lambda-\nu}{2}}-\frac{dp}{dr}=0. \label{tov}
\end{equation}

Here $M_G=M_G(r)$ is the effective gravitational mass within the sphere up to  radius $r$ which takes form
\begin{equation}
M_G(r)=\frac{1}{2}r^2e^{\frac{\nu-\lambda}{2}}\nu^{\prime}.
\label{egm}
\end{equation}

The above eq. (\ref{tov}),  can be expressed as
\begin{equation}
 F_g+ F_h=0,\label{Force}
\end{equation}
where
\begin{equation}
F_{g}=-\frac{2}{3}\rho\left(\frac{g_{tt}^{\prime}}{g_{tt}}\right),
\end{equation}\label{eq34}

\begin{equation}
F_{h}=-\frac{\rho^{\prime}}{3}.
\end{equation}\label{eq36}

The above eq. (\ref{tov}) or eq. (\ref{Force}), therefore indicates that subject to the gravitational and hydrostatic forces the equilibrium is achieved for the fluid sphere.

\subsection{The Homotopy Perturbation Method (HPM)}
Primarily our task is to find the solution of eq. (\ref{eq9}) by using the HPM. Therefore to have a basic idea about HPM,
let us consider the following nonlinear differential equation
\begin{equation}
A(u) - f(r)=0,\label{eq9a}
\end{equation}
under the boundary condition
\begin{equation}
B\left(u,\frac{\partial u}{\partial n}\right)=0.\label{eq9b}
\end{equation}

Here we identify $A$ as a general differential operator, $B$ as boundary operator, $f(r)$ as a known analytical function and $\frac{\partial}{\partial n}$ as directional derivative. The operator $A$ is divided into linear and nonlinear part, namely $L$ and $N$. So eq. (\ref{eq9a}) takes the form
\begin{equation}
L(u) + N(u) - f(r)= 0. \label{eq9c}
\end{equation}

Now a homotopy is constructed as
\begin{equation}
H(v,p)= L(v) -L(u_{0})+ h L(u_{0})+h[N(v)- f(r)],
\end{equation}
where $h$ is an embedding parameter and $u_{0}$ is an initial approximation. As $h$ changes from 0 to 1 monotonically, $H(v,0)$ is continuously transformed into $H(v,1)$. In topology this is called continuous deformation and hence $H(v,0)$ and $H(v,1)$ are homotopic functions.

Under the above formalism, now to solve eq. (\ref{eq9}) we consider linear part of the differential equation as $m^{\prime} = dm/dr$ and the rest of the terms as non-linear. So the homotopy equation can be written as~\citep{He2000}
\begin{equation}
\hspace{-0.3cm} m^{\prime} - m_0^{\prime} + h
\left[m_0^{\prime}+mm^{\prime\prime} -
\frac{1}{2}m^{\prime\prime}r-\frac{2}{3}(m^{\prime})^{2}-\frac{4mm^{\prime}}{r}\right]
=0,\label{eq10}
\end{equation}
where $m_0$ is the initial guess of the
solution. For $h=0$, we get initial approximation solution and for
$h=1$, one obtains the desired solution for mass.

We assume the initial solution as $m_{0}=ar^{3}$. According to
HPM, therefore, we consider the  general solution as
\begin{equation}
m = m_{0}+hm_{1}+h^2m_{2}+h^3m_{3}+....\label{eq11}
\end{equation}

Substituting this in eq. (\ref{eq10}) and equating coefficients of
the different orders of $h$, we get the solution for mass up to second
order correction as follows
\begin{equation}
m(r) = ar^{3} +
\frac{36}{5}a^{2}r^{5}+\frac{312}{35}a^{3}r^{7}+3Aar^{2}+B,\label{eq12}
\end{equation}
where $A$ and $B$ are constants of integration.

The boundary conditions in the present system are
\begin{equation}
m(0)=0,~m^{\prime}(R)=0.\label{eq13}
\end{equation}

Using eqs. (\ref{eq12}) and (\ref{eq13}) we obtain the constants of integration, $A$ and $B$ in the following form
\begin{equation}
A = - \left(\frac{R}{2} + 6aR^{3}+\frac{52}{5}a^{2}R^{5}\right),\label{eq14}
\end{equation}

\begin{equation}
B=0,\label{eq16}
\end{equation}
where $R$ is the radius of the spherical distribution.

Therefore, substituting $A$ in eq. (\ref{eq12}), we obtain
\begin{equation}
 m(r)=ar^{3}+\frac{36}{5}a^{2}r^{5}+\frac{312}{35}a^{3}r^{7}-3ar^{2}\left(\frac{R}{2}+6aR^{3}+\frac{52}{5}a^{2}R^{5}\right).\label{eq17}
\end{equation}

\section{The Einstein field equations and their solutions}
The Einstein field equations for our system are given by
\begin{equation}
\frac{2m^{\prime}}{r^{2}}= 8\pi\rho, \label{EFE1}
\end{equation}

\begin{equation}
\frac{2m}{r^{3}} -\left(1-\frac{2m}{r}\right)\frac{g_{tt}^{\prime}}{g_{tt}}\frac{1}{r}= -8\,{\pi}p, \label{EFE2}
\end{equation}

\begin{eqnarray}
-\left(1-\frac{2m}{r}\right)\left[\frac{1}{2}\frac{g_{tt}^{\prime\prime}}{g_{tt}}
-\frac{1}{4}\left(\frac{g_{tt}^{\prime}}{g_{tt}}\right)^{2} +
\frac{1}{2r}\frac{g_{tt}^{\prime}}{g_{tt}}\right]  \nonumber
\\ -\left (\frac{m}{r^{2}}-\frac{m^{\prime}}{r}\right)
\left(\frac{1}{r} +\frac{1}{2}\frac{g_{tt}^{\prime}}{g_{tt}}\right)= -8\,{\pi}p.
\end{eqnarray}

Now, substituting eq. (\ref{eq17}) into eq. (\ref{EFE1}) we obtain an expression for density as follows
\begin{equation}
\rho = \frac{1}{4\pi}\left(3a+36a^{2}r^{2}+\frac{312}{5}a^{3}r^{4}\right)- \frac{3a}{2\pi r}\left(\frac{R}{2}+6aR^{3}+\frac{52}{5}a^{2}R^{5}\right),\label{eq22}
\end{equation}
whereas, using eq. (\ref{eq22}) in eq. (\ref{eq2}), we get an expression for the radial pressure as
\begin{equation}
p=\frac{1}{12\pi}\left[3a+36a^{2}r^{2}+\frac{312}{5}a^{3}r^{4} -\frac{6a}{r}\left(\frac{R}{2}+6aR^{3}+\frac{52}{5}a^{2}R^{5}\right)\right].\label{eq30a}
\end{equation}

Again by using eq. (\ref{eq2}) and eq. (\ref{EFE1}), we solve eq. (\ref{EFE2})
for $g_{tt}$ as follows:
\begin{equation}
g_{tt} = \frac{K\exp{\int\frac{4}{3(r-2m)}dr}}{r(r-2m)^{1/3}},\label{eq23}
\end{equation}
where $K$ is a constant.

\begin{figure}[h]
\centering
\includegraphics[width=6cm]{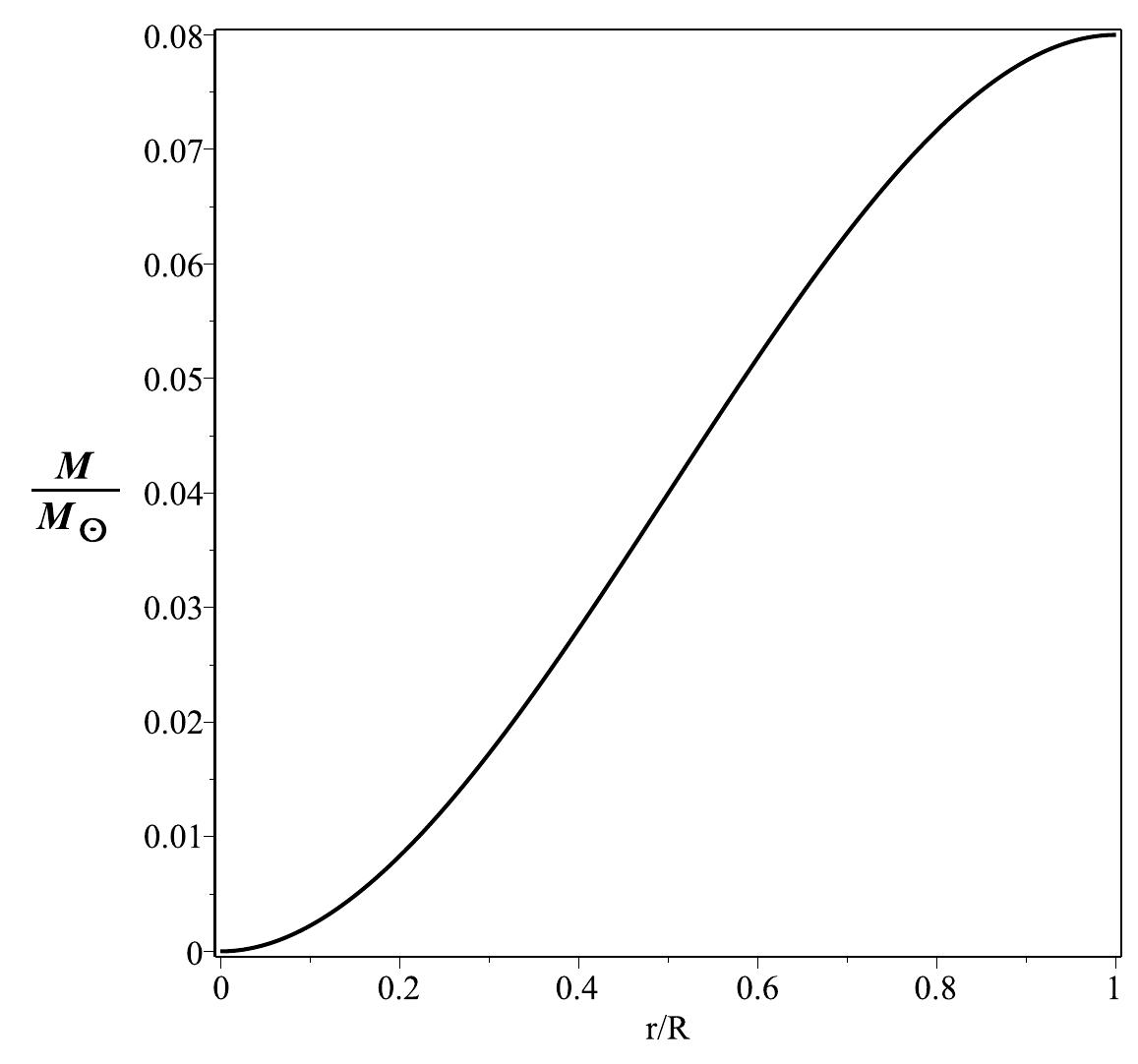}
\caption{Variation of the mass as a function of the fractional radial distance (r/R) for $a= -2.55 \times
10^{-15}~{{km}^{-2}}$ and $R=0.065~{R_{\odot}}$   } \label{fig1}
\end{figure}

\begin{figure}[h]
\centering
\includegraphics[width=6cm]{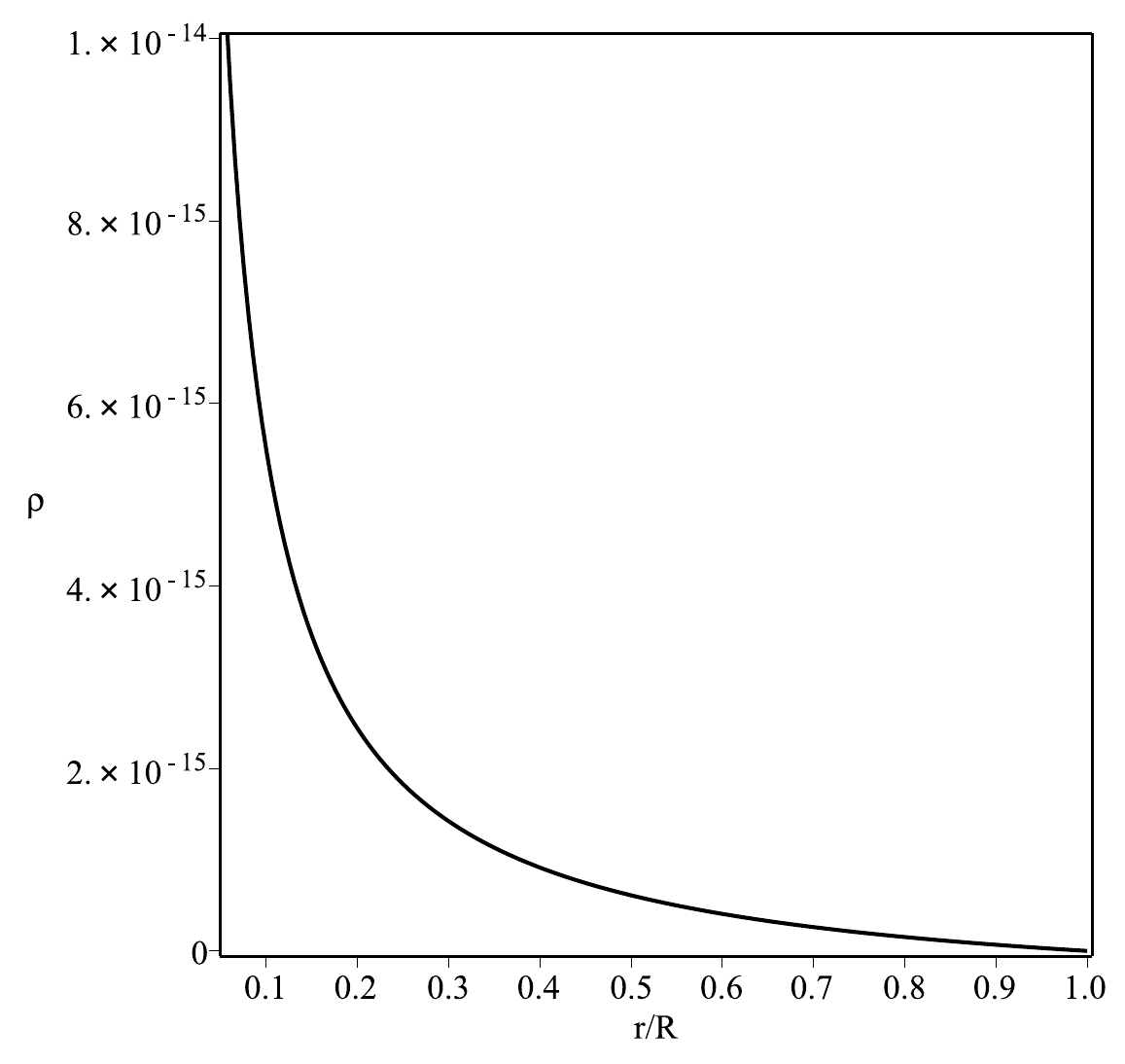}
\caption{Variation of the density as a function of the fractional radial distance (r/R) for $a= -2.55 \times
10^{-15}~{{km}^{-2}}$ and $R=0.065~{R_{\odot}}$  } \label{fig2}
\end{figure}

\begin{figure}[h]
\centering
\includegraphics[width=6cm]{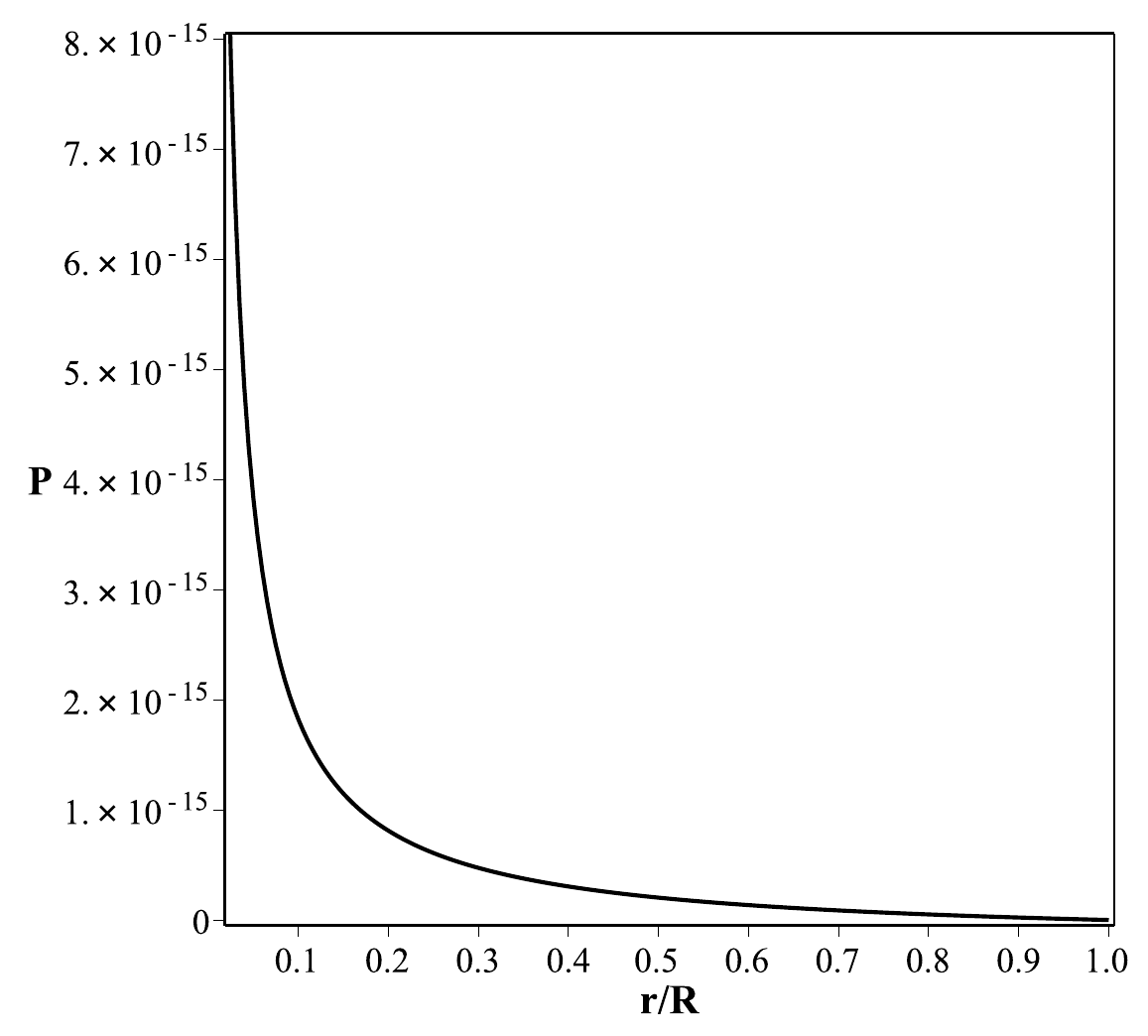}
\caption{Variation of the radial pressure as a function of the fractional radial distance (r/R)  for $a= -2.55 \times
10^{-15}~{{km}^{-2}}$ and $R=0.065~{R_{\odot}}$  } \label{fig3}
\end{figure}


\section{Physical features of the radiation model}

\subsection{Mass, density and pressure}
To start with the basic physical studies preferably we choose
parameter $a= -2.55 \times {{10}^{-15}} {{km}^{-2}}$ and $R=0.065$ solar radius
and $m(R)=0.080$ solar mass of $E0$ class of brown dwarf stars
(see the Link~\citet{StarTables}). The reason to consider these
data set lies on the physical background that our target is to
investigate a radiating model which is compatible with brown
dwarfs which are cool star-like objects that have insufficient
mass to maintain stable nuclear fusion in their core regions
~\citep{Rebolo1995}. It has been argued~\citet{Rebolo1995} that
although brown dwarfs are not stars, they are expected to form in
the same way, and they should radiate a large fraction of their
gravitational energy at near-infra red wavelengths.

The variation of mass, density and pressure with respect to the radial coordinate
is shown in figures~\ref{fig1}-\ref{fig3}. 

\subsection{Stability of stellar model}

\begin{figure}
\centering
\includegraphics[width=6cm]{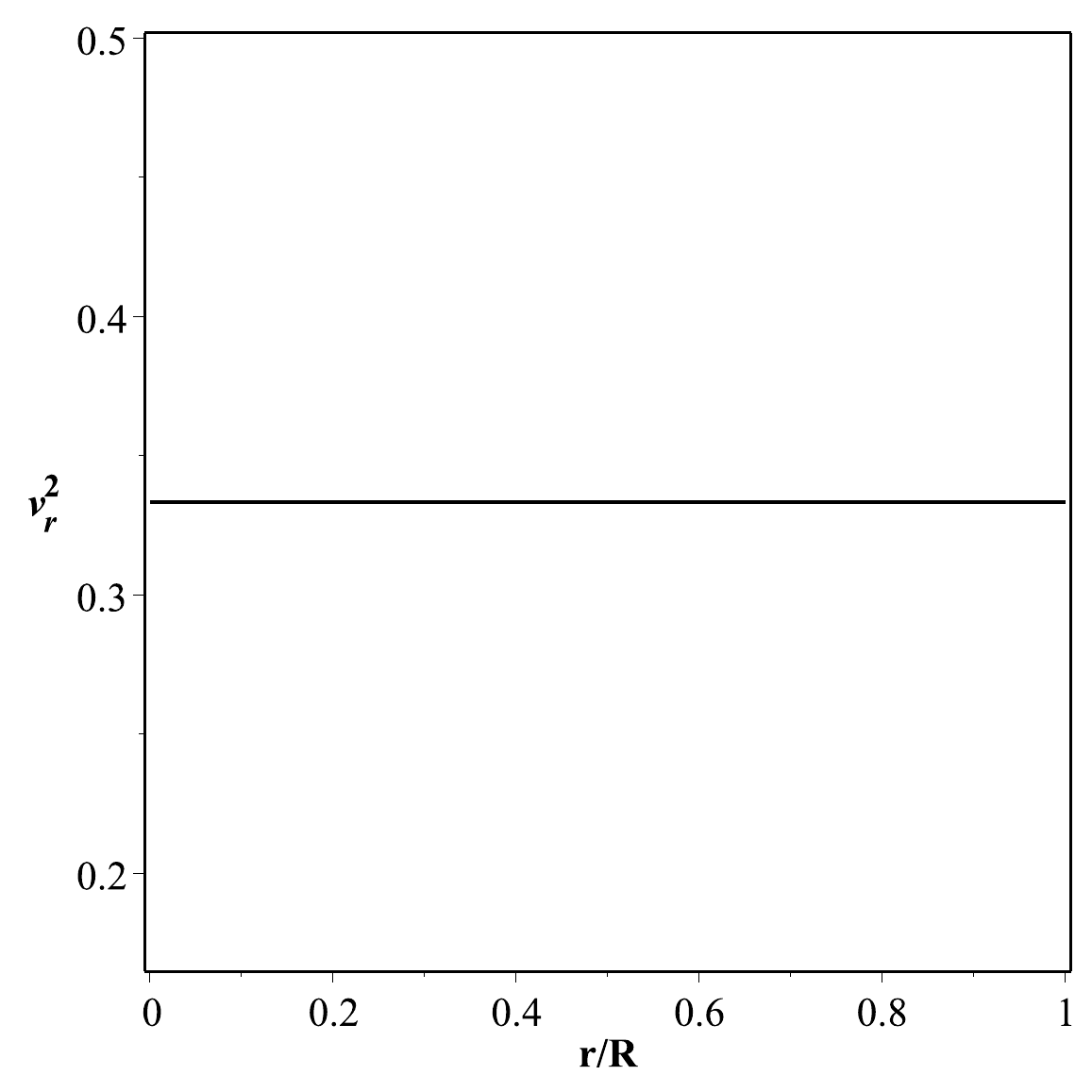}
\caption{Variation of the forces as a function of the fractional radial distance (r/R)  for $a= -2.55 \times
10^{-15}~{{km}^{-2}}$ and $R=0.065~{R_{\odot}}$  } \label{fig10}
\end{figure}

Now we perform some physical test regarding the stability of stellar model. First we check the sound speed in the interior of stellar structure. The sound speed is defined as
\begin{equation}
v^{2}_{r}= \frac{dp}{d\rho}.
\end{equation}
For a physical perfect fluid model square of sound speed must be less than the speed of light in the interior of star \citep{Herrera}. This leads to the condition $v^{2}_{r}<1$.
Now from figure~\ref{fig10} we see that  in our model the square of sound speed has constant value which is $1/3$.

\begin{figure}
\centering
\includegraphics[width=6cm]{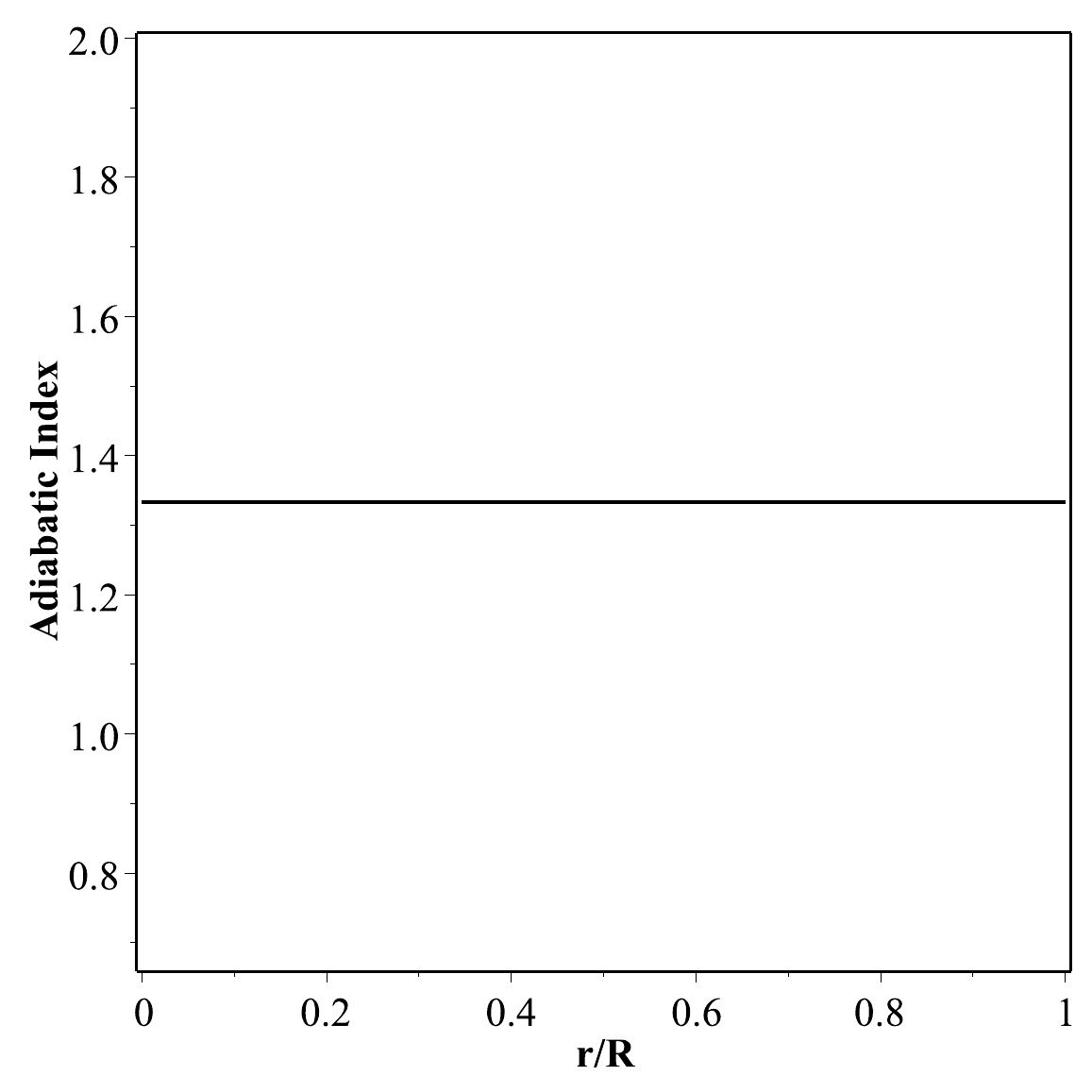}
\caption{Variation of the forces as a function of the fractional radial distance (r/R) for $a= -2.55 \times
10^{-15}~{{km}^{-2}}$ and $R=0.065~{R_{\odot}}$  } \label{fig11}
\end{figure}

The adiabatic index for a perfect fluid structure is defined by
\begin{equation}
\Gamma=\left(\frac{\rho+p}{p}\right) \left[\frac{dp}{d\rho}\right].
\end{equation}
The value of adiabatic index will be greater than $4/3$ for a stable configuration at equilibrium. From figure~\ref{fig11} we find that adiabatic index is equal to $4/3$  which is the critical value of adiabatic index~\citep{Chandrashekhar1964,Bondi1964,Wald1984,Das2016}.

\subsection{Energy conditions}
Let us now check whether all the energy conditions regarding present radiating model are satisfied or not. For this purpose, we
are considering the following inequalities and plotting graphs for each case:\\
(i)~NEC: for~any~null~vector~$K^{\alpha},~ T_{\alpha\beta}K^{\alpha}K^{\beta} \geq 0,$
or~effectively~$\rho+p\geq 0.$\\
(ii)~WEC: for~any~timelike~vector~$V^{\alpha},~ T_{\alpha\beta}V^{\alpha}V^{\beta} \geq 0,$
or~effectively~$\rho \geq 0$~and~$\rho+p\geq 0.$\\
(iii)~SEC: for~any~timelike~vector~$V^{\alpha},~ (T_{\alpha\beta}-\frac{1}{2}Tg_{\alpha\beta})V^{\alpha}V^{\beta} \geq 0,$
or~effectively~$\rho+p\geq 0$~and~$\rho+3\,p\geq 0.$\\
(iv)~DEC: for~any~timelike~vector~$V^{\alpha},$\\
(a)~$T_{\alpha\beta}V^{\alpha}V^{\beta} \geq 0$~and~$T_{n}^{\alpha}V^{n}$~is~causal,\\
(b)~for~any~ two~co-oriented~timelike~vectors~$V^{\alpha}$~and~$W^{\alpha},\\~T_{\alpha\beta}V^{\alpha}W^{\beta} \geq 0$
~or~effectively~$\rho \geq 0$~and~$\rho> \mid p \mid.$

\begin{figure}
\centering
\includegraphics[width=6cm]{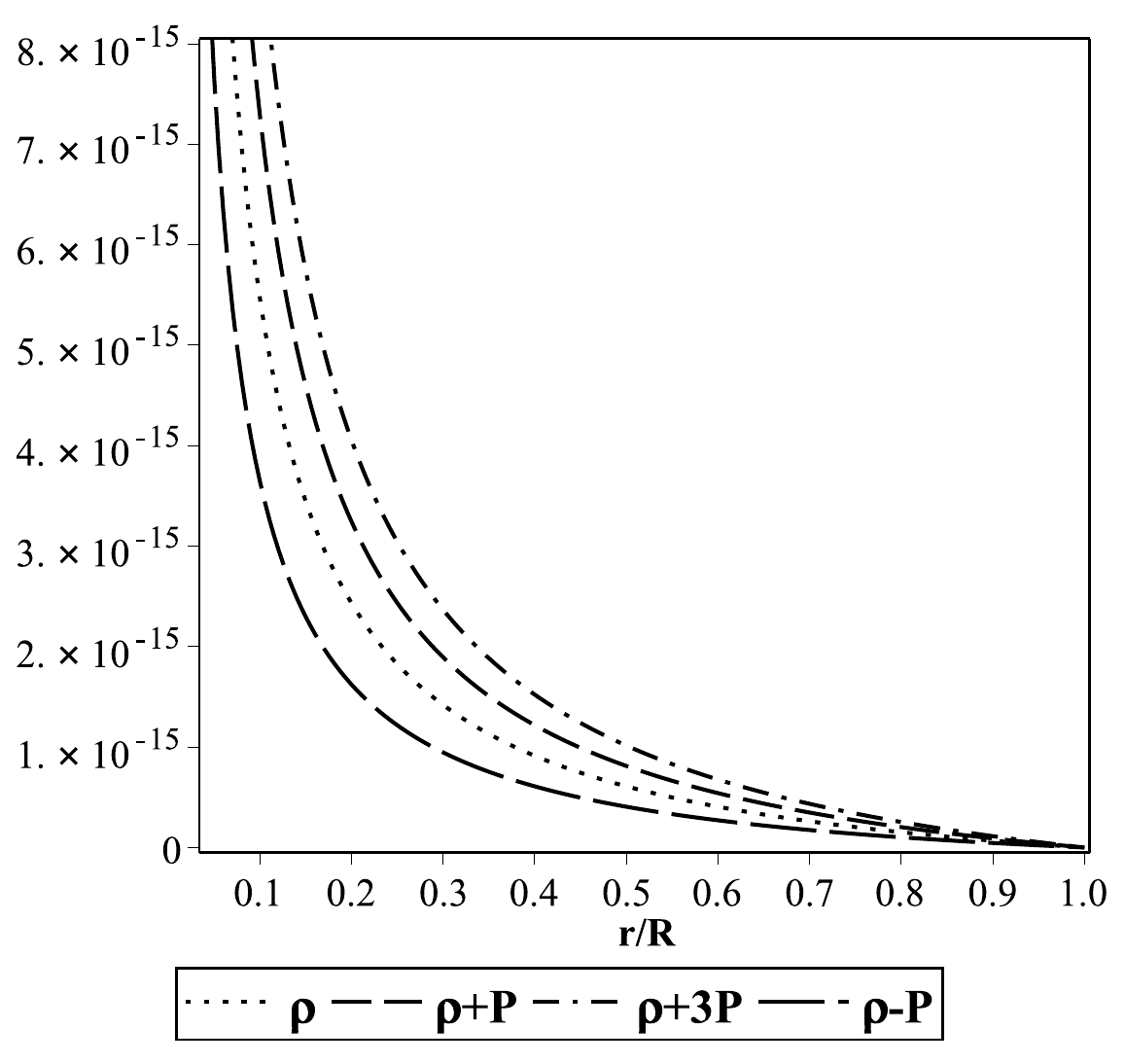}
\caption{The energy conditions  applicable in the interior of the
spherical distribution are plotted against $r/R$ for $a= -2.55 \times
10^{-15}~{{km}^{-2}}$ and $R=0.065~{R_{\odot}}$  }\label{fig5}
\end{figure}

We note from figure~\ref{fig5}, that in our model all the energy conditions are satisfied throughout the interior region.

\subsection{Compactness and redshift}
By definition compactness of a star can be given by $u=m(r)/r$ which in our case takes the following form:
\begin{equation}
u=ar^{2}+\frac{36}{5}a^{2}r^{4}+\frac{312}{35}a^{3}r^{6}
-3ar\left(\frac{R}{2}+6aR^{3}+\frac{52}{5}a^{2}R^{5}\right).\label{eq27}
\end{equation}

\begin{figure}
\centering
\includegraphics[width=6cm]{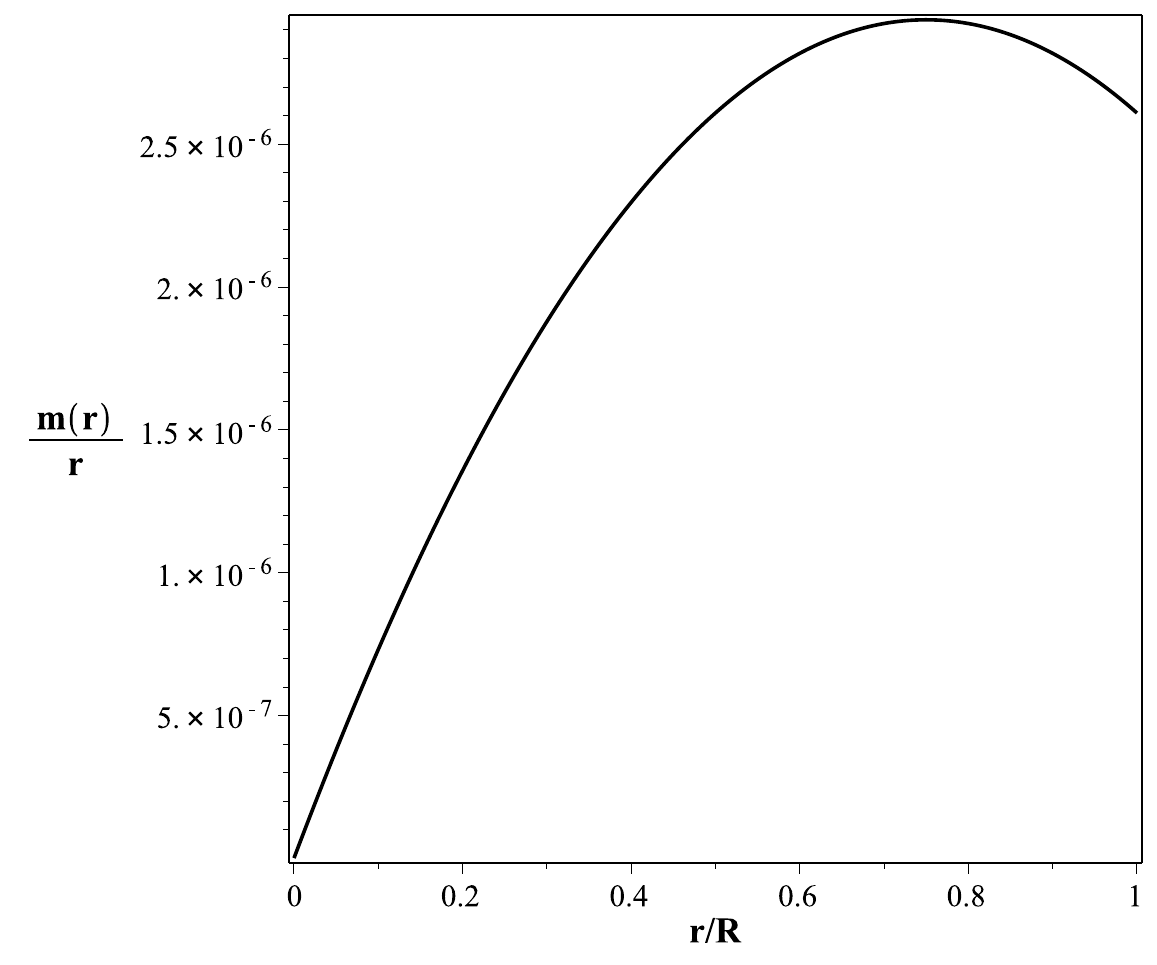}
\caption{Variation of the compactification factor as a function of the fractional radial distance for $a= -2.55 \times
10^{-15}~{{km}^{-2}}$ and $R=0.065~{R_{\odot}}$  }\label{fig6}
\end{figure}

It can easily be observed that figure~\ref{fig6} exhibits the condition $\frac{m(R)}{R}< \frac{4}{9}$. 

Here the value of $u$ is of the order of $10^{-6}$ in relativistic unit throughout the star. It can be noted that the exponential term in eq. \ref{eq23} becomes $r^{\frac{4}{3}}$ under the approximation $2u\leq 1$ and hence we obtain 
\begin{equation}
g_{tt} \approx  \frac{Kr^{1/3}}{(r-2m)^{1/3}}. \label{eq24}
\end{equation}

Using boundary conditions for $g_{tt}$, one can find out $K$,
which ultimately gives the time-time component of metric (\ref{eq1}) as
\begin{equation}
g_{tt} =
\frac{\left[1-\frac{2m(R)}{R}\right]^{4/3}}{\left[1-\frac{2m(r)}{r}\right]^{1/3}},
\label{eq25}
\end{equation}
where $m(R)$ is a constant and can be given by
\begin{equation}
m(R) =
-\frac{1}{2}aR^{3}-\frac{54}{5}a^{2}R^{5}-\frac{156}{7}a^{3}R^{7}.
\label{eq26}
\end{equation} 

Again, the surface redshift is defined by the relation
\begin{equation}
z= (1-2u)^{-\frac{1}{2}} -1,\label{eq28}
\end{equation}
which in the present study turns out to be
\begin{equation}
1+z= \frac{1}{\sqrt{1-2
\{ar^{2}+\frac{36}{5}a^{2}r^{4}+\frac{312}{35}a^{3}r^{6}-3ar\left(\frac{R}{2}+6aR^{3}+\frac{52}{5}a^{2}R^{5}\right)\}}}.
\label{eq29}
\end{equation}

\begin{figure}
\centering
\includegraphics[width=6cm]{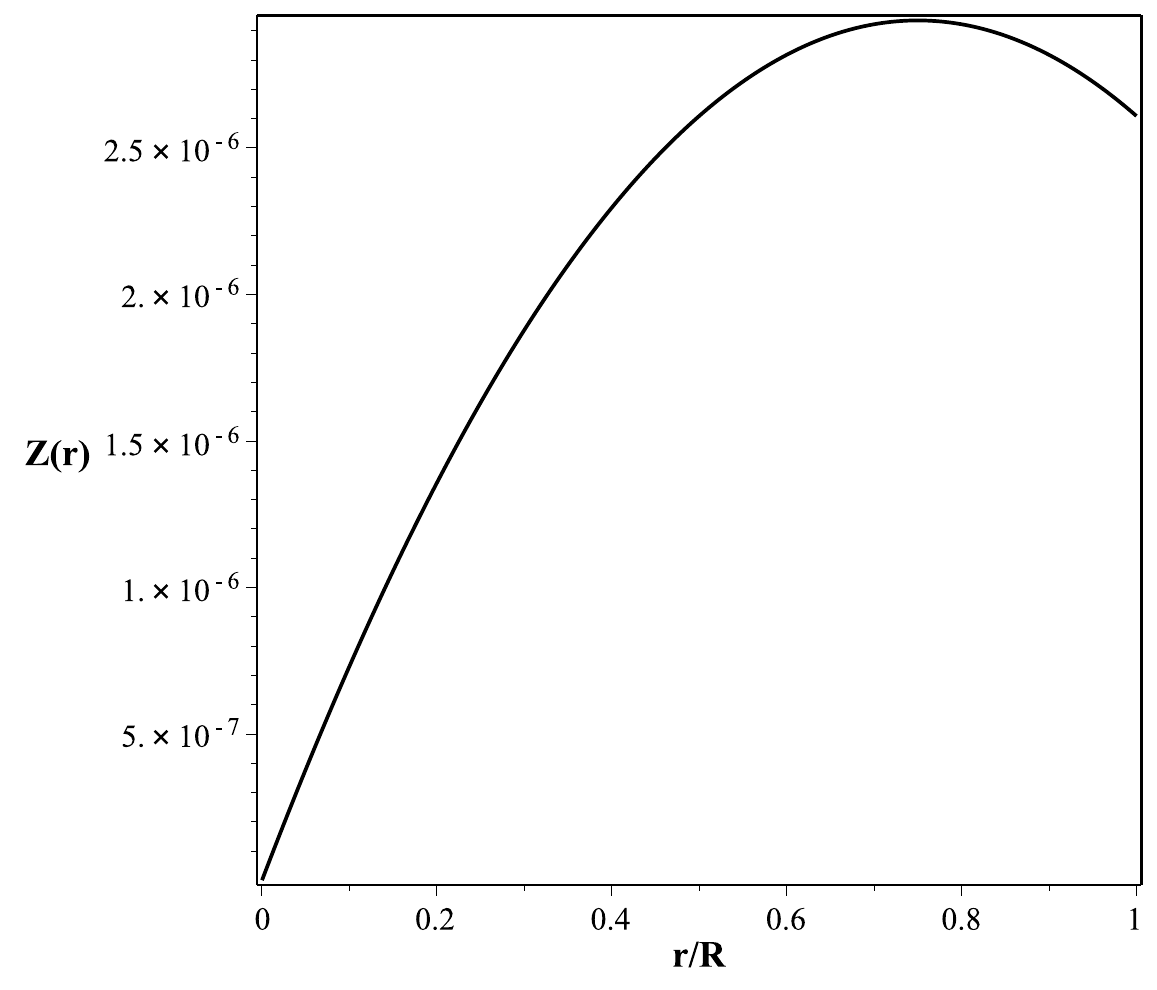}
\caption{Variation of the redshift as a function of the fractional radial distance for $a= -2.55 \times
10^{-15}~{{km}^{-2}}$ and $R=0.065~{R_{\odot}}$  }\label{fig7}
\end{figure}

The variation of surface redshift $z$ is shown in  figure~\ref{fig7} which is physically within an acceptable profile.

\section{Conclusion}
Basically the goal of the present paper is to study a radiating star on the basis of electromagnetic radiation by exploiting the equation
of state $p = {\frac{1}{3}}\rho$. For this purpose our first step is involved in finding out an expression of mass for a spherically symmetric system from the Euler-Lagrangian equation by using Homotopy Perturbation Method. Hence by employing this expression for the mass and the Einstein field equations we have obtained a set of interior solution. We have explained different physical properties of the solutions describing the system. In the present model for radiating brown dwarf stars all the features are seen to be physically viable.

 One may find a few articles where HPM have been used in the field of astrophysics and the trend is getting it's way successfully~\citep{Shchigolev2013,	Shchigolev1,Shchigolev2,Shchigolev3,Shchigolev4}. We have applied HPM using particular homotopy structure which gives a particular solution describing brown dwarf stars.

In connection to all the above aspects with brown dwarf stars we have done a preliminary calculation taking the data set of~\citet{Bhar2015} related to highly compact stars such as neutron stars and strange stars. This data set provides negative
radial pressure and also suffers from instability. Therefore, we suspect that radiation model may not be compatible with highly compact stars rather it is compatible with radiating brown dwarf stars. Actually, to relate radiation model with highly compact stars is also not justified due to their exhausted fuel system and hence burning as well as radiation stage which seems have been stopped much earlier.

Regarding singularity we note that there are several famous solutions available in the literature \citep{Tolman1939,Zeldovich1962,Misner1964} which suffer from this unwanted situation by exhibiting infinite curvature at the centre. In this connection we would also like to point out that according to an analysis by Delgaty and Lake \citep{DL1998} out of 127 published solutions only 16 solutions satisfy all the physical conditions of general relativity. However, from eq. \ref{eq25} of the present model we have $g_{tt}(0)=0.999993$ as finite and positive whereas $g_{rr}(0)=1$. But the density and pressure are not finite at the centre though we have finite mass. So the solution is geometrically non-singular with infinite density and pressure at centre of the star. This type of solution physically possible as discussed in their work by Fuloria and Durgapal \citep{Fuloria2008}. They argued that there is a possibility of having information from the central region of infinite density and pressure if the central redshift is finite. In our model for the brown dwarf the redshift in the central region of the star is finite. Inspired by this fact we can think of a core at the central region of the star. The central density for a brown dwarf varies from $10$ $gm/cm^{3}$ to $10^{3}$ $gm/cm^{3}$ \citep{Burrows1993}. From figure \ref{fig2}  we see that the density becomes $10^{3}$ $gm/cm^{3}$ at certain critical distance  ($r_{c}=369$ $km$) near the centre. Since the star reaches its maximum density at $r_{c}$, we consider a small core of constant density $10^{3}$ $gm/cm^{3}$  of core radius $r_{c}$. This is a good approximation as the core is very small compared to the star (the volume of the core is just $0.000054311$ percent of the total volume of the star). Therefore we find the constant density core as suitable alternative to the centre of infinite density.

 In future work it may be possible to get a non-singular solution with finite central density and pressure using different homotopy structure. As a final remark we would like to state that though in the present radiation model the brown dwarf star has been successfully described yet some more investigations are needed to perform with different methodology before coming to a definite decision regarding applicability of HPM and MEP to radiating and highly compact stars.

\section*{Acknowledgments} FR and SR are thankful to the Inter-University
Centre for Astronomy and Astrophysics (IUCAA), Pune, India for providing the
Visiting Associateship under which a part of this work was carried
out. SR also thanks the Institute of Mathematical Sciences (IMSc), Chennai, India 
for providing the working facilities and hospitality under the Associateship scheme.

\end{document}